\documentclass[journal=jacsat,manuscript=article]{achemso}

\usepackage[T1]{fontenc} 
\usepackage{amsmath}
\usepackage{amsfonts}
\usepackage{amsthm}
\usepackage{mathtools}
\usepackage{xcolor}  
\usepackage{wrapfig}
\usepackage{float}
\usepackage{subcaption}
\usepackage{graphicx}

\newcommand{\uf}{{\bf f}}

\newcommand{\ux}{{\bf x}}

\newcommand{\uwt}{U_{\rm wt}}
\newcommand{\umu}{U_{\rm mut}}
\newcommand{\rd}{{\rm d}}

\usepackage[%
  colorlinks=true,
  allcolors=blue
]{hyperref}

\usepackage{natbib}
\usepackage{textcomp}

\newcommand{\rfig}[1]{Figure~\ref{#1}}

\newcommand{\req}[1]{eq~\ref{#1}}

\usepackage[version=3]{mhchem} 

\author{Yajie Cai}
\altaffiliation{These authors contributed equally.}
\affiliation[Purdue University]{Department of Chemistry, Purdue University, West Lafayette, IN 47907}

\author{Yanbin Wang}
\altaffiliation{These authors contributed equally.}
\affiliation[Purdue University]{Department of Chemistry, Purdue University, West Lafayette, IN 47907}

\author{Ming Chen}
\affiliation[Purdue University]{Department of Chemistry, Purdue University, West Lafayette, IN 47907}
\email{chen4116@purdue.edu}

\title{DeltaDiff: Training-Free, Physics-Guided Machine Learning for Predicting Mutant Protein Structures}

\abbreviations{IR,NMR,UV}
\keywords{American Chemical Society, \LaTeX}
\SectionNumbersOn

\begin{document}

\begin{tocentry}
    \centering
    \includegraphics[width=3.25in,height=1.75in,keepaspectratio]{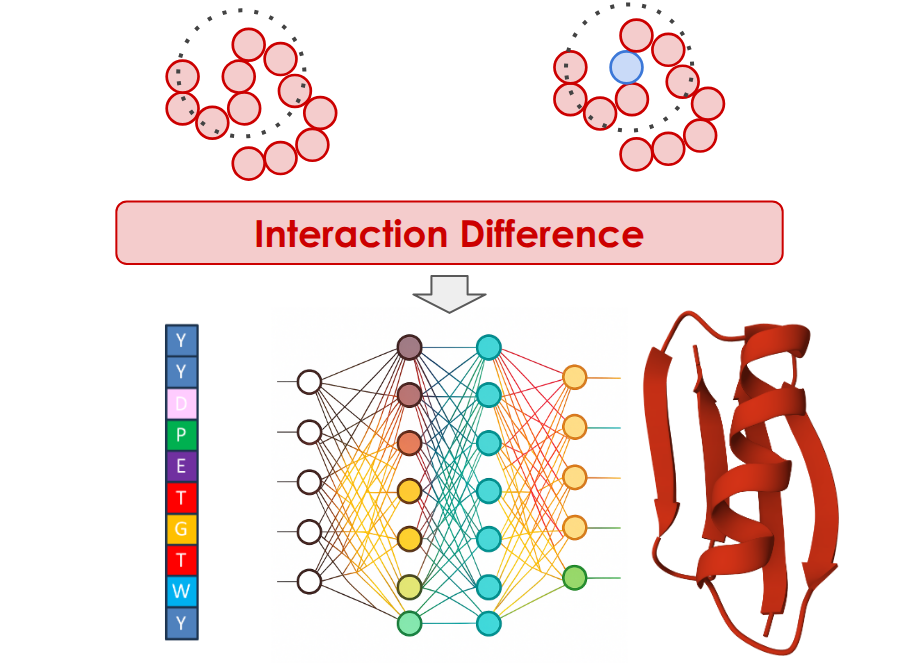}

\end{tocentry}

\begin{abstract}
Determining mutant protein structures is critical for understanding the mechanistic roles of mutations in biochemical processes. However, experimental characterization and conventional theoretical modeling are often expensive and time-consuming. Recent advances in machine learning provide new opportunities to efficiently predict protein structures from primary sequences. Nevertheless, applying these models to proteins with single-site or few-site mutations remains challenging because mutant sequences are often highly similar to their wild-type counterparts.
Here, we introduce DeltaDiff, a physics-guided inference framework for mutant-structure generation that incorporates mutation-aware physical guidance into a baseline diffusion model. We evaluate DeltaDiff on three representative systems: Chignolin T8P, Novispirin G-10, and BBL D162N. All three examples involve nonlocal structural changes, making accurate mutant-structure prediction challenging.
DeltaDiff captures key mutation-induced conformational changes without requiring retraining or fine-tuning of the baseline model. These results establish a foundation for efficient mutant-structure prediction at a fraction of the cost of conventional methods, facilitating rational mutant design.

\end{abstract}

\section{Introduction}
Protein mutation studies are central to biochemistry, with broad applications in identifying functionally important residues for biofunctions, drug development, and protein engineering \cite{lek2016analysis, ebrahimi2023engineering, depristo2005missense, soskine2010mutational, tsuboyama2023megascale, karczewski2020mutational}. Modeling how mutations reshape protein structure provides mechanistic insight into biochemical processes such as drug resistance \cite{blair2014molecular,darby2022molecular}, pathology-related protein-protein interactions \cite{sahni2015widespread}, and provides rational design rules for protein engineering \cite{tokuriki2009stability, worth2009structural, cheng2021comprehensive}.

However, understanding mutant protein structures remains challenging. Although many mutations cause only local structural perturbations, such as changes in side-chain packing, numerous examples show that even a single mutation can substantially alter protein structure \cite{solomon2023reversible, he2012mutational}. 
The mutation effects can propagate through local packing, hydrogen-bond networks, and allosteric pathways \cite{pace2014contribution, bigman2018stability, naganathan2019modulation}.
For instance, mutations can induce loop restructuring, which may strongly affect enzyme activity \cite{corbella2023loop}. In metamorphic proteins, an alternative misfolded state can have stability comparable to that of the native folded state, and even a single mutation can invert their relative stability and trigger significant secondary-structure changes \cite{tuinstra2008interconversion, bryan2010proteins,solomon2023reversible, he2012mutational}.

Conventionally, mutation effect can be studied through both experimental and computational approaches. Experimental methods, such as deep mutational scanning or phage display, provide high-throughput approaches for screening large mutant libraries \cite{fowler2014deep, fowler2010highresolution, smith1985filamentous, alexander2005directed}. However, many proteins are not good targets for high-throughput mutation assays if their real function cannot be coupled to a scalable readout \cite{araya2011deep}. Moreover, most high-throughput approaches cannot resolve mutant conformations, which are crucial for the mechanistic understanding of the correlation between mutations and protein functions.  Experimental techniques that determine three-dimensional protein structures, including nuclear magnetic resonance (NMR) \cite{wuthrich1989protein, wthrich2001way}, cryo-electron microscopy (cryo-EM) \cite{khlbrandt2014resolution,nakane2020single}, and X-ray crystallography \cite{kendrew1958three, perutz1960structure}, are expensive and time-consuming when applied to large mutant libraries \cite{ii2008understanding, chayen2008protein}. 
Conventional computational methods often rely on molecular dynamics (MD) simulations to determine mutant protein structures \cite{hou2023single, bromley2020tumorigenic, patel2021replica, han2012exploring, galdadas2020unravelling, kotzampasi2024free, lilkova2012metadynamics}. While MD simulations are powerful for modeling local structural changes, determining mutant structures with nonlocal structural changes with MD is time-consuming\cite{shaw2010atomic,lindorff2011fast}.

Recent advances in machine learning (ML) have enabled a low-cost, high-throughput method for generating structures directly from amino-acid sequences \cite{baek2021accurate, mirdita2022colabfold,jumper2021highly, evans2021protein, rives2021biological, lin2023evolutionary,lewis2025scalable,abramson2024accurate,watson2023novo, lu2024str2str}. 
For example, protein language models, such as AlphaFold2\cite{jumper2021highly} and ESM\cite{lin2023evolutionary}, achieve highly accurate structure prediction for many proteins.
Besides predicting a single structure, other generative models aim to predict protein structural ensembles. For example, score-based diffusion models (SBDM), such as RFDiffusion\cite{watson2023novo} Str2Str\cite{lu2024str2str}, and BioEmu\cite{lewis2025scalable}, use stochastic processes to transform a simple Gaussian distribution into a distribution of the protein structural ensemble.

However, predicting the structures of mutant proteins with ML remains less reliable. Experimentally relevant mutations often involve only one or a few residues, which poses a fundamental challenge for current AI models: small sequence perturbations may produce only weak changes in the model input or internal representations, while still causing substantial physical consequences. For example, the success of AlphaFold2 relies heavily on multi-sequence alignment\cite{yang2023alphafold, jumper2021highly,rao2021msa}. When a mutant sequence differs from the wild type by only one residue, its MSA representation may remain highly similar to that of the wild type. As a result, the predicted mutant structure may remain artificially close to the wild-type structure \cite{buel2022alphafold, pak2023using,mcbride2023alphafold,agarwal2024power}. Although more recent models, such as AlphaFold3 \cite{abramson2024accurate}, have improved capabilities for modeling protein mutant structures, their performance remains case-dependent\cite{wee2024evaluation}, as demonstrated in this work. Accurate prediction of mutant protein structures with ML requires models to learn mutation-induced structural responses from large and diverse datasets. However, experimentally resolved mutant structures remain limited and biased toward stable, crystallizable proteins, making it difficult for models to capture nonlocal rearrangements and conformational shifts. Therefore, new ML-based sampling methods are needed to address this challenge.  
One promising strategy is to guide the inference of a pretrained ML model using physically meaningful, mutation-specific interactions. 
Such guidance can improve the ability of foundation models to generate mutant conformations. 

In our previous work, we demonstrated that physical guidance can steer a pretrained SBDM towards protein structures that are either consistent with experimentally measured features \cite{liu2025exendiff} or responsive to changes in molecular environment \cite{wang2025extrapolating}. 
Here, we extend this physically guided inference strategy to mutant-structure generation. By incorporating a coarse-grained (CG) potential energy function as a guidance potential during inference \cite{majewski2023machine}
, we developed an SBDM inference method, called DeltaDiff, for mutation-aware conformational sampling 
The CG potential energy function quantifies mutation-induced changes in residue-level interactions between mutant and wild-type sequences, and this interaction difference is used to bias the SBDM inference. 
The manuscript will be organized as follows. First, we will present the theory of DeltaDiff. Second, DeltaDiff will be evaluated on three single-mutation systems that involve distinct structural responses: Chignolin T8P, Novispirin G-10, and BBL D162N. Across these cases, DeltaDiff enriches mutant-like conformations that are weakly sampled by the unguided baseline, suggesting that physics-guided inference provides a practical route for adapting pretrained generative models to mutation-induced structural changes without retraining on large mutant-structure datasets.

\section{METHODS}
\subsection{A Brief introduction to score-based diffusion model}

We used a pretrained SBDM as the foundation generative model for protein backbone conformation sampling. An SBDM defines a continuous noising process that gradually perturbs data samples into a simple prior distribution, followed by a learned reverse process that transforms noisy samples back into structured data. Let $\mathbf{x}$ denote the protein structural variables, which may represent Cartesian coordinates, residue frame translations, or other continuous structural degrees of freedom. The forward diffusion process can be written as the stochastic differential equation (SDE)
\begin{equation}
    \rd \mathbf{x}
    =
    \uf(\mathbf{x},t)\,\rd t
    +
    g(t)\,\rd \mathbf{w},
    \qquad t\in[0,1],
    \label{eq:forward_sde_mut_guidance}
\end{equation}
where $\uf(\mathbf{x},t)$ is the drift term, $g(t)$ is the diffusion coefficient, and $\mathbf{w}$ is a standard Wiener process. As $t$ increases, the data distribution is progressively changed to a white-noise distribution that is easier to sample. 

To generate a new sample that follows the data distribution, SBDM uses a reverse-time SDE\cite{song2020scorebased, anderson1982reverse}: 
\begin{equation}
    \rd \mathbf{x}
    =
    \left[
    \uf(\mathbf{x},t)
    -
    g^2(t)\nabla_{\mathbf{x}}\log P(\mathbf{x},t)
    \right]\rd t
    +
    g(t)\rd \bar{\mathbf{w}},
    \label{eq:reverse_sde_mut_guidance}
\end{equation}
where $\bar{\mathbf{w}}$ is a reverse-time Wiener process and $P(\mathbf{x},t)$ is the marginal distribution at diffusion time $t$. The term $\nabla_{\mathbf{x}}\log p_t(\mathbf{x})$, known as the ``score function''. In practice, the score function is modeled by a neural network $\mathbf{s}_{\boldsymbol{\theta}}(\mathbf{x}(t),t)$ that is parametrized by $\boldsymbol{\theta}$, allowing the reverse process to iteratively denoise random coordinates into plausible protein conformations.

\subsection{Theory of DeltaDiff}
We assume that there is a pretrained SBDM that generates wild-type protein structures using residue-level CG representations following the Boltzmann distribution $P_{\mathrm{SBDM}}\approx P_{\mathrm{wt}}(\mathbf{x})\propto e^{-\beta \uwt(\ux)}$, where $\mathbf{x}$ is the CG coordinates of the protein and $\uwt(\ux)$ is an effective residue-level energy function implicitly encoded in the SBDM. $\beta=1/k_{\rm B}T$ where $k_{\rm B}$ is the Boltzmann constant and $T$ is temperature. Although training SBDMs to quantitatively reproduce the Boltzmann distributions of diverse proteins remains a challenging task, recent advances in SBDMs provide increasingly accurate approximations of protein conformational distributions \cite{lee2023scorebased, lewis2025scalable, ho2020denoising, wang2026conditional}. For a mutant, we can write the structural probability distribution as $P_{\mathrm{mut}}(\mathbf{x})\propto e^{-\beta \umu(\ux)}$. Since $\ux$ is a residue-level CG representation, changing the type of a residue does not alter the dimensionality of $\ux$. Instead, the mutation is represented by changing the ``atom type'', which transforms $\uwt(\ux)$ into $\umu(\ux)$.
To generate mutant conformations, we bias the SBDM towards $P_{\mathrm{mut}}$ during reverse diffusion with a correction: 
\begin{equation}
    P_{\mathrm{mut}}(\mathbf{x}) \propto P_{\mathrm{SBDM}}(\mathbf{x}) \exp[-\beta \Delta U_{\mathrm{mut-wt}}(\mathbf{x})].
    \label{eq:target_mut_distribution}
\end{equation}
where \(\Delta U_{\mathrm{mut-wt}}(\mathbf{x})=\umu(\ux)-\uwt(\ux)\) is the mutation-induced energy difference.

One way to incorporate $\Delta U_{\mathrm{mut-wt}}$ into SBDM inference is to introduce a time-dependent guidance potential:
\begin{equation}
     \Delta U_t(\mathbf{x}(t))
    =
    -k_{\mathrm{B}}T
    \log
    \mathbb{E}_{p(\mathbf{x}(0)|\mathbf{x}(t))}
    \left[
    \exp[-\beta \Delta U_{\mathrm{mut-wt}}(\mathbf{x}(0))]
    \right].
    \label{eq:time_dependent_mutation_potential}
\end{equation}
where $p(\mathbf{x}(0)|\mathbf{x}(t))$ is the conditional distribution of clean structures $\mathbf{x}(0)$ generated from a given noisy configuration $\mathbf{x}(t)$ by the original SBDM. 
With Eq.(\ref{eq:time_dependent_mutation_potential}), the reverse diffusion process
\begin{equation}
        \rd \mathbf{x}
    =
    \left[
    \mathbf{f}(\mathbf{x},t)
    -
    g^2(t)
    \left(
    \mathbf{s}_{\boldsymbol{\theta}}(\mathbf{x}(t),t)
    +
    \nabla_{\ux(t)}\Delta U_{t}(\mathbf{x}(t),t)
    \right)
    \right]\rd t
    +
    g(t)\rd\bar{\mathbf{w}}
    \label{eq:mut_guided_reverse_sde}
\end{equation}
can sample from $P_{\mathrm{mut}}(\mathbf{x})$ \cite{song2020scorebased}. However, evaluating the expectation over $p(\mathbf{x}(0)|\mathbf{x}(t))$ requires repeatedly sampling the conditional denoising distribution at each diffusion time $t$, which is computationally expensive. To reduce this cost, we approximate the conditional distribution by a delta distribution centered at the denoised estimate:
\begin{equation}
    P(\mathbf{x}(0)|\mathbf{x}(t))
    \approx
    \delta(\mathbf{x}(0)-\hat{\mathbf{x}}(0))\;,
    \label{eq:denoised_estimate_approx}
\end{equation}
where $\hat{\mathbf{x}}(0)=\mathbb{E}(\ux(0)|\ux(t))$. 
Under this approximation, the time-dependent guidance potential simplifies to
\begin{equation}
    \Delta U_t(\mathbf{x}(t))
    \approx
    \Delta U_{\mathrm{mut-wt}}(\hat{\mathbf{x}}(0)).
    \label{eq:delta_u_xhat}
\end{equation}
Therefore, $P_{\mathrm{mut}}(\mathbf{x})$ can be approximately sampled using the guided reverse diffusion process
\begin{equation}
    \rd \mathbf{x}
    =
    \left[
    \mathbf{f}(\mathbf{x},t)
    -
    g^2(t)
    \left(
    \mathbf{s}_{\boldsymbol{\theta}}(\mathbf{x}(t),t)
    +
    \lambda(t)
    \widetilde{\mathbf{F}}_{\mathrm{mut-wt}}(\mathbf{x}(t),t)
    \right)
    \right]\rd t
    +
    g(t)\rd\bar{\mathbf{w}}\;,
    \label{eq:mut_guided_reverse_sde}
\end{equation}
where $\widetilde{\mathbf{F}}_{\mathrm{mut-wt}}(\mathbf{x}(t),t)=\nabla_{\ux(t)}\Delta U_{\mathrm{mut-wt}}(\hat{\mathbf{x}}(0))$. Here, $\lambda(t)$ is an empirical time-dependent scaling factor, which will be discussed below.

In this work, we use Str2Str as the baseline model to define ($P_{\mathrm{SBDM}}$), as it has been successfully applied in other guided-inference studies. One advantage of Str2Str is that it uses a reference structure and can control the similarity between the generated and reference structures. Using the wild-type structure as a reference, this feature allows us to selectively focus on either localized structural perturbations or more global conformational changes in mutant proteins.
We use a residue-level many-body CG model \cite{majewski2023machine} to define $\uwt$ and $\umu$. Although this model was trained on a limited number of proteins, our results suggest that it can effectively guide Str2Str to sample mutant structures.

\begin{figure}[H]
    \centering
    \includegraphics[width=6.5in]{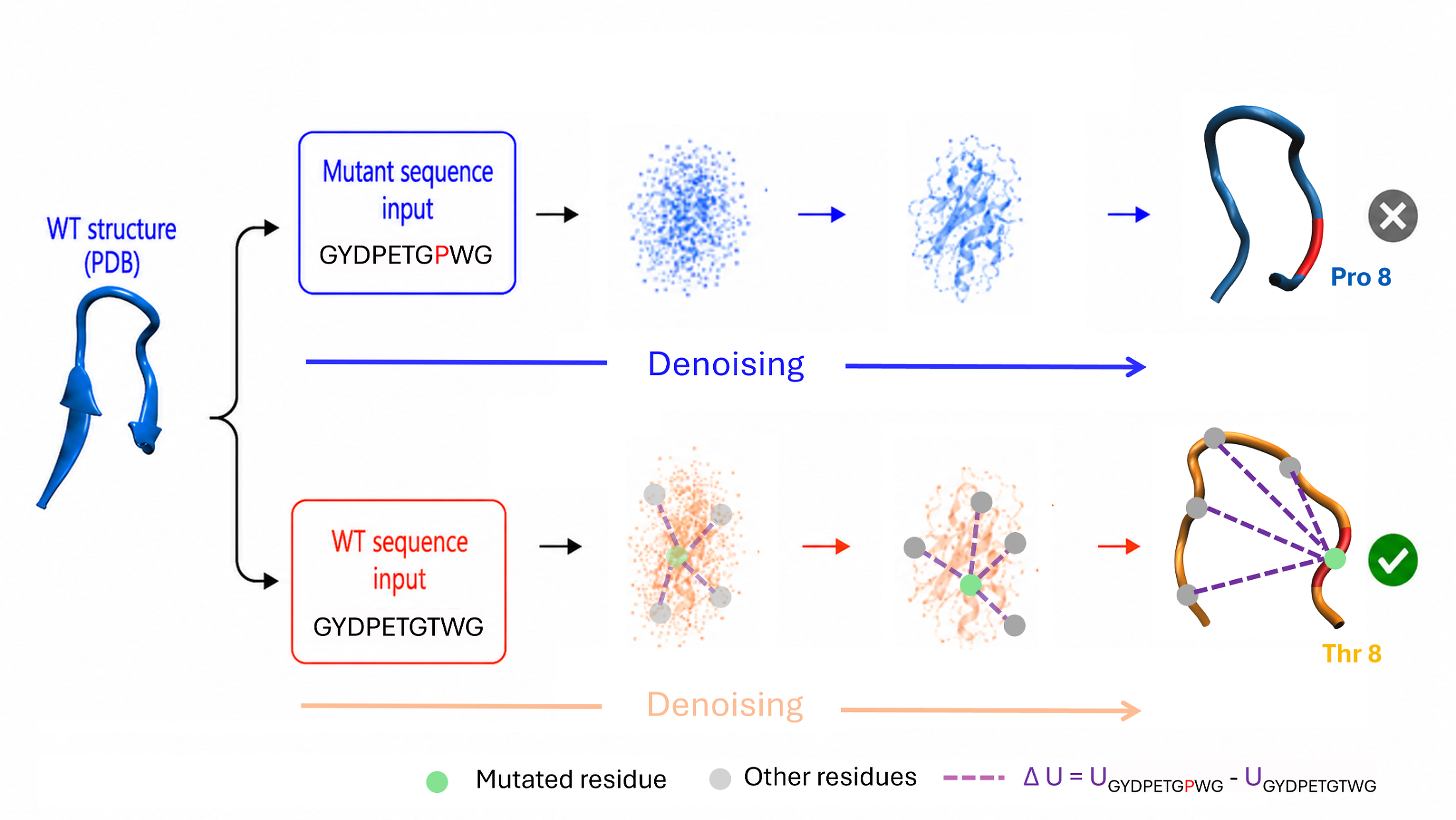}
    \caption{Schematic illustration of mutant-structure generation with physically guided diffusion models. Traditional ML takes the blue inference pathways: direct structure generation conditioned on the mutant sequence. DeltaDiff uses the orange inference pathway: physics-guided inference. In the second pathway, the SBDM is guided by the interaction energy difference between the wild-type and mutant sequences, $\Delta U = U_{\rm mut} - U_{\rm wt}$, illustrated by the purple dashed lines. Direct mutant-sequence generation tends to produce conformations that remain close to the wild-type structure. In contrast, guided sampling allows the generated structures to explore mutant conformations.}
    \label{fig:schematic}
\end{figure}

Since both Str2Str and the CG model represent protein backbone conformations at the residue level, side-chain packing must be reconstructed using backmapping methods. We used AttnPacker \cite{mcpartlon2023endtoend} with the \textsc{AttnPackerPTM\_V2} model to backmap the coarse-grained representations to all-atom structures. The resulting packed structures were then relaxed using PyRosetta FastRelax \cite{chaudhury2010pyrosetta} with the ref2015 score \cite{alford2017rosetta}, three FastRelax repeats, and coordinate constraints to the starting structures to allow gentle local relaxation. The summary of the DeltaDiff algorithm is illustrated in \rfig{fig:schematic}.

\section{RESULTS}
In this section, we present case studies to assess the ability of DeltaDiff to capture key conformational changes induced by single-site mutations. 
We examine three representative systems: Chignolin T8P, a 10-residue peptide; Novispirin G-10, an 18-residue peptide; and BBL D162N, a 50-residue protein. These systems represent single-site mutations that induce distinct conformational responses. The DeltaDiff-generated structures are validated against experimental measurements for Novispirin G-10 and against simulation-based references for chignolin T8P and BBL D162N. Additional validation details are provided in the Supporting Information.
\subsection{Chignolin with mutation of T8P}
\begin{figure}[H]
    \centering
    \includegraphics[width=6in]{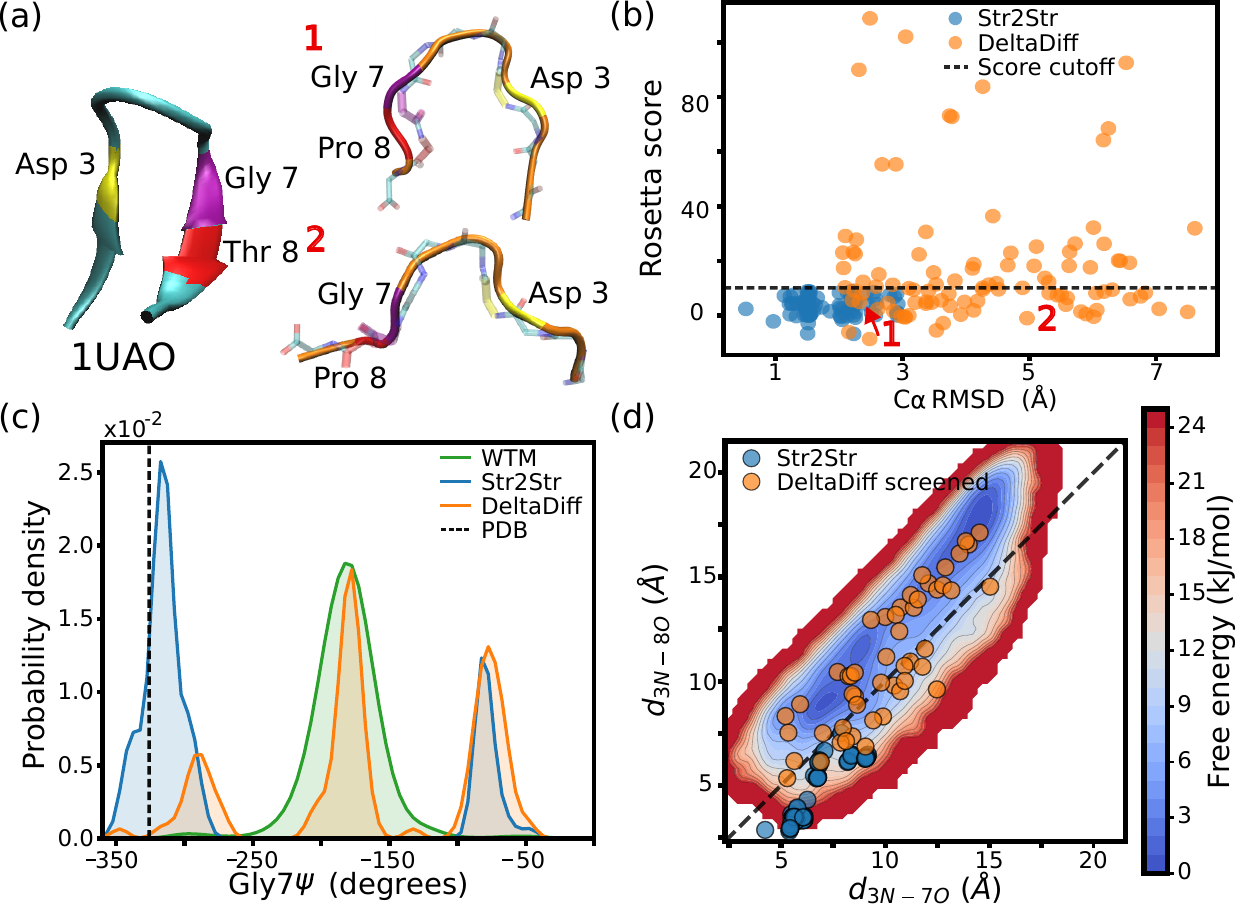}
    \caption{{\bf(a)} Folded state of chignolin structure (PDB ID: 1UAO) and two representative T8P mutant structures generated by DeltaDiff. Asp3, Gly7, and Thr8/Pro8 are labeled, with Thr8/Pro8 indicating the mutation site. The two mutant structures are overlaid with their closest MD-sampled structures in the background. {\bf(b)} Rosetta relaxed T8P structures are projected onto Rosetta score and C$\alpha$ RMSD to the wild type PDB structure. The dashed line marks the median DeltaDiff relaxed-score cutoff used for screening; representative structures 1 and 2 are shown in Panel~(a). {\bf(c)} Gly7 $\psi$ dihedral angle distributions are presented. The black dotted line marks the 1UAO value. The green/blue/orange curve indicates the distribution from well-tempered metadynamics simulations/Str2Str/DeltaDiff {\bf(d)} Generated structures are projected onto the MD free energy surface defined by the amide N of Asp3 to Gly7 carbonyl O distance ($d_{\rm 3N-7O}$) and the amide N of Asp3 to Pro8 carbonyl O distance ($d_{\rm 3N-8O}$). The dashed diagonal line marks $d_{\rm 3N-7O}=d_{\rm 3N-8O}$. Conformations above the line favor the $\alpha$-turn.}
    \label{fig:chignolin}
\end{figure}

Chignolin ~\cite{honda200410} (GYDPETGTWG) is a widely used benchmark system in both experimental and computational studies to understand protein folding \cite{maruyama2020mutation, satoh2006chignolin}. 
Previous experimental and computational studies have shown that the wild-type Chignolin (Chignolin WT) folds into a $\beta$-hairpin structure with a $\pi$-turn (residues Asp3-Thr8) \cite{maruyama2020mutation}. Due to the backbone constraints introduced by proline, Chignolin T8P forms an $\alpha$-turn (residues Asp3-Gly7). We performed well-tempered metadynamics \cite{barducci2008welltempered} simulations on Chignolin T8P, and the sampling results also suggest that $\alpha$-turn is the dominant conformation of Chignolin T8P (\rfig{fig:chignolin}(a)).

We generated 100 Chignolin T8P structures using the baseline Str2Str model and another 100 structures using DeltaDiff. All structures from Str2Str and DeltaDiff are relaxed with Rosetta. 
We then used the Rosetta scores of the relaxed structures to rank the generated conformations and identify low-score configurations that are more likely to represent physically plausible low-free-energy states of Chignolin T8P. 
To characterize the generated structures, we projected them onto two features: (1) the $\alpha$-carbon ($C_{\alpha}$) RMSD relative to the folded structure of Chignolin WT and (2) the Rosetta scores of the relaxed structures, as shown in \rfig{fig:chignolin}(b).
For subsequent structural analysis, we retained the low-score structures generated by DeltaDiff.

We further validated the structural ensemble generated by DeltaDiff using well-tempered metadynamics. Following previous studies, we used the Gly7 $\psi$ dihedral angle as a descriptor (\rfig{fig:chignolin}(c)) \cite{maruyama2020mutation}. For Chignolin WT, $\psi \approx -330^\circ$, while well-tempered metadynamics suggests that $\psi$ fluctuates between $-150^\circ$ and $-250^\circ$ for Chignolin T8P. The Chignolin T8P structures predicted by Str2Str are mainly distributed around $\psi \approx -330^\circ$, while DeltaDiff significantly increases the probability of sampling mutant conformations with the correct $\psi$ torsions.
In addition to the Gly7 $\psi$ dihedral angle, we selected two distances that are directly associated with the $\pi$-to-$\alpha$ turn transition: the distance between the Asp3 amide N atom and the Gly7 carbonyl O atom, denoted as 3N-7O ($d_{\rm 3N-7O}$), and the distance between the Asp3 amide N atom and the Pro8 carbonyl O atom, denoted as 3N-8O ($d_{\rm 3N-8O}$). For Chignolin WT, $d_{\rm 3N-7O}>d_{\rm 3N-8O}$ because the structure forms a $\pi$-turn, while for Chignolin T8P, $d_{\rm 3N-7O}<d_{\rm 3N-8O}$ because the mutant forms a $\alpha$-turn.
As shown in \rfig{fig:chignolin}(d), most DeltaDiff-generated structures satisfy $d_{\rm 3N-7O}<d_{\rm 3N-8O}$, consistent with the $\alpha$-turn conformation of Chignolin T8P. In contrast, Str2Str predicts Chignolin T8P structures that remain similar to the wild-type structure, with $d_{\rm 3N-7O}>d_{\rm 3N-8O}$.

\subsection{G10 NOVISPIRIN with mutation of I10G of Ovispirin-1}
Ovispirin-1 (KNLRRIIRKIIHIIKKYG, PDB: 1HU5) is an antimicrobial peptide with high cytotoxicity \cite{steinstraesser2002activity}. The mutant peptide Novispirin G-10 (PDB:1HU6) replaces Ile10 with Gly, which reduces the undesired toxicity of Ovispirin-1 \cite{sawai2002impact}. 

We used the same strategy as in the Chignolin T8P case to sample and analyze mutant structures: DeltaDiff-generated structures were screened using the Rosetta score, and low-score structures were selected for further analysis. We first mapped the generated structures onto two features: (1) the $\alpha$-carbon ($C_{\alpha}$) RMSD relative to Novispirin G-10 and (2) the Rosetta scores of the relaxed structures, as shown in \rfig{fig:1HU6}(a). Str2Str-generated configurations generally exhibit $C_{\alpha}$ RMSD values greater than 3.5~\AA, indicating that Str2Str rarely samples structures close to the experimentally resolved mutant conformation. In contrast, DeltaDiff-generated configurations show a broader distribution along the $C_{\alpha}$ RMSD coordinate and cover lower-RMSD regions, reaching approximately 2.5~\AA. 

To further analyze the DeltaDiff-generated structures, we examined a key structural feature of the mutant: the helical backbone becomes noticeably curved. We defined a bending angle using two vectors, one connecting the N-terminus to Gly10 and the other connecting Gly10 to the C-terminus, as illustrated in the inset of \rfig{fig:1HU6}(b). The I10G mutation likely weakens the side-chain interactions that constrain the wild-type peptide in a relatively straight helical conformation. {In agreement with this interpretation, the bending angle of the wild-type structure is close to $30^\circ$ (see Supporting Information), whereas that of the experimentally resolved mutant structure 1HU6 is close to $80^\circ$.
We then compared the bending-angle distributions of the experimental structural ensemble, Str2Str-generated structures, and DeltaDiff-generated structures. The DeltaDiff ensemble samples a broader distribution with bending angles approaching $\sim60^\circ$. In contrast, Str2Str-generated structures remain concentrated blow $30^\circ$, which is even smaller than the bending angle of the wild-type structure (\rfig{fig:1HU6}(b)). This comparison further supports that DeltaDiff can generate mutant structures that capture the key structural transformation induced by the I10G mutation.

\begin{figure}[H]
    \centering
    \includegraphics[width=6in]{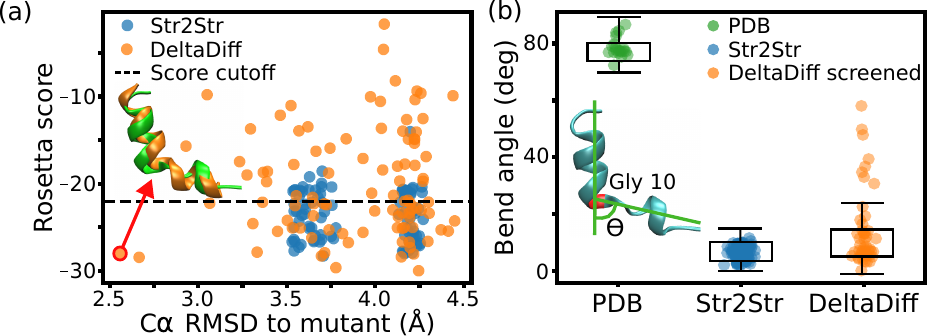}
    \caption{{\bf(a)} Relaxed I10G structures are projected onto the two-dimensional feature space defined by Rosetta score and C$\alpha$ RMSD to the mutant 1HU6 reference structure. The dashed line marks the median DeltaDiff relaxed-score cutoff used for screening. The inset shows a representative DeltaDiff-generated structure with the lowest RMSD to 1HU6 and a low Rosetta score (orange), overlaid with the 1HU6 structure (green).
{\bf(b)} Bend-angle distributions of the experimental and generated I10G structures. The inset illustrates the bend-angle definition, computed as the angle between two C$\alpha$-based vectors: one from the N-terminal segment to Gly10 and the other from Gly10 to the C-terminal segment.}
    \label{fig:1HU6}
\end{figure}

\subsection{BBL with mutation of D162N}
BBL (GSQNNDALSPAIRRLLAEWNLDASAIKGTGVGGRLTREDVEKHLAKA PDB:2WXC), a small peripheral subunit-binding domain capable of cooperative ultrafast folding, provides a representative model for studying protein folding mechanisms \cite{neuweiler2009folding, cho2008origins, yu2008cooperative}. 
Previous experiments reported that the D162N mutant significantly reduces the melting temperature \cite{neuweiler2009folding}, highlighting the substantial effect of this mutation on BBL stability.

We used the same screening strategy for the BBL D162N mutant, retaining DeltaDiff-generated structures with low Rosetta scores. To characterize mutation-induced conformational changes, we projected the sampled structures onto two features: (1) the Rosetta score and (2) the Ala130-Arg157 C$\alpha$-C$\alpha$ distance ($d_{\rm 130C_\alpha-157C_\alpha}$). Since the D162N mutation perturbs the local hydrogen-bonding and electrostatic interactions between Asp162 and Thr152/159, this perturbation leads to restructuring of the loop linking two helices (residues 143-159) and further changes the interactions between this loop and the loop at the N-terminus (residues 124-132), resulting in a large $d_{\rm 130C_\alpha-157C_\alpha}$ (\rfig{fig:bbl}(a)). We performed 2~$\mu$s MD simulations on BBL D162N starting from a wild-type-like structure. The loop conformation in the wild-type-like structure is not stable and quickly relaxes to a more stable structure with large $d_{\rm 130C_\alpha-157C_\alpha}$. We emphasize that the 2~$\mu$s MD simulation is not sufficient to exhaustively sample unfolded states of BBL D162N. Therefore, our analysis focuses on folded conformations and mutation-induced structural rearrangements within the folded-state ensemble.
As shown in \rfig{fig:bbl}(b), the structures generated by Str2Str are concentrated in a narrow region with a small $d_{\rm 130C_\alpha-157C_\alpha}<1$ nm and centered near $d_{\rm 130C_\alpha-157C_\alpha}<5$ \AA, corresponding to the wild-type BBL. In contrast, DeltaDiff-generated structures cover a much broader range of distances, extending up to approximately 2~nm, which indicates that DeltaDiff explores conformations beyond the wild-type-like one.

Next, we compared the DeltaDiff structures with MD-sampled D162N mutant conformations. The two helical segments, residues 133-142 and 160-170, remain stable throughout the simulation, while the major mutation-induced conformational variability arises mainly from rearrangements of two loops.
To characterize the sampled conformations, we projected the MD structures onto two structural features: (1) $d_{\rm 130C_\alpha-157C_\alpha}$ and (2) the distance between the Gly153 backbone H atom and the Arg157 backbone O atom, denoted as 153H--157O ($d_{\rm 153H-157O}$), which indicates a stable backbone hydrogen bond in the helix-linking loop. 
We constructed a free energy surface by projecting the MD configurations onto the feature space. The structures generated by Str2Str and DeltaDiff were also projected onto the same feature space for comparison (\rfig{fig:bbl}(c)).
Both Str2Str and DeltaDiff generate many structures with large $d_{\rm 153H-157O}$ that cannot form a hydrogen bond. This is possibly due to the limited ability of Str2Str to predict low-free-energy loop conformations in BBL. Interestingly, within the limited number of Str2Str-predicted structures, all have a short $d_{\rm 130C_\alpha-157C_\alpha}$, making the structures similar to the wild type BBL. We also tested using AlphaFold3 to predict BBL D162N structures. Unfortunately, AlphaFold3 can only generate wild-type-like structures (see Supporting Information). 
In contrast, DeltaDiff samples a broader region of the feature space and enhances the possibility of exploring the low-free-energy conformation with $d_{\rm 130C_\alpha-157C_\alpha}\approx 10$~\AA and $d_{\rm 153H-157O}\approx 2$~\AA. 
Notably, the Str2Str explores conformations with high free energies, according to the free energy surface in \rfig{fig:bbl}(c), but some Str2Str-generated structures have low Rosetta scores. This observation suggests that the Rosetta score can serve as a useful tool for screening generated structures, but identifying the correct mutant conformation requires a more accurate energy function.

\begin{figure}[H]
    \centering
    \includegraphics[width=6in]{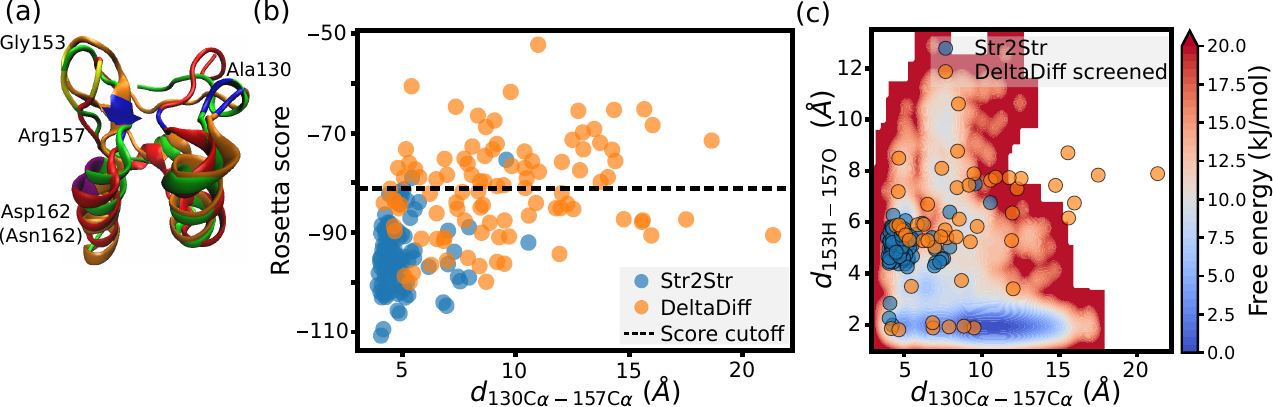}
    \caption{{\bf(a)} Conformational change of the BBL D162N. The wild-type structure (PDB ID: 2WXC) is shown in red, the dominant D162N MD structure in green, and a representative DeltaDiff-generated D162N structure in orange. Key residues Ala130 and Arg157 are highlighted in blue, Gly153 is highlighted in yellow, and the mutation site Asp162/Asn162 is highlighted in purple. {\bf(b)} Relaxed D162N structures projected onto the two-dimensional feature space defined by Rosetta score and the 130-157 C$\alpha$ distance ($d_{\rm 130C_\alpha-157C_\alpha}$). The dashed line marks the median DeltaDiff relaxed-score cutoff used for screening. {\bf(c)} Generated structures projected onto the MD free energy surface defined by the $d_{\rm 130C_\alpha-157C_\alpha}$ and the distance between the Gly153 backbone H atom and the Arg157 backbone O atom ($d_{\rm 153H-157O}$).}
    \label{fig:bbl}
\end{figure}

\section{DISCUSSION}
The practical implementation of DeltaDiff requires an approximation to the time-dependent effective interaction during reverse diffusion (see \req{eq:mut_guided_reverse_sde}). This approximation is reliable near \(t\approx 0\), where the noisy structure \(\mathbf{x}(t)\) remains close to the denoised structure \(\mathbf{x}(0)\), but becomes less accurate near \(t\approx 1\), where \(\mathbf{x}(t)\) contains little information about \(\mathbf{x}(0)\).
The construction of \(\lambda(t)\) is therefore empirical, but its asymptotic behavior is physically motivated. As \(t\rightarrow 1\), the conditional distribution \(p(\mathbf{x}(0)|\mathbf{x}(t))\) becomes weakly dependent on \(\mathbf{x}(t)\). In this limit, the exact effective potential is approximately independent of \(\mathbf{x}(t)\), and the corresponding guidance force should vanish. By contrast, as \(t\rightarrow 0\), the approximation of \(\mathbf{x}(0)\) from \(\mathbf{x}(t)\) becomes more accurate, and the physical guidance can be applied at full strength.
Accordingly, we introduce a time-dependent scaling factor and apply the guided drift as \(\lambda(t)\mathbf{F}(\mathbf{x},t)\), with
$\lambda(t)\rightarrow 0 \quad \text{as } t\rightarrow 1,
\qquad
\lambda(t)\rightarrow 1 \quad \text{as } t\rightarrow 0 .$
This scaling suppresses the guidance force in the early noisy regime and gradually activates it during the late denoising regime. While the specific functional form of \(\lambda(t)\) is chosen empirically, any smooth schedule satisfying these limiting conditions can be used. In our previous work, we tested multiple functional forms for this purpose and found that the detailed choice of the schedule had only a minor effect on sampling performance \cite{e27030291}. Therefore, in DeltaDiff, we adopt one representative $\lambda(t)$ for all applications considered in this study: a sigmoid switching function,
\begin{equation}
\lambda(t) = \frac{1}{1 + e^{ -\tau \left( t_{\mathrm{mid}} - t \right)} }, \quad t_{\mathrm{mid}} = 0.5
\label{eq:force_scale_sigmoid}
\end{equation}
where $t_{\mathrm{mid}}$ controls the diffusion time at which the guidance is activated most rapidly, and $\tau$ controls the sharpness of the transition. 
The choice of a sigmoid function is motivated by recent observations in SBDMs: the reverse diffusion process exhibits a sharp transition from noise-like coordinates at large $t$ to structure-like coordinates at small $t$. 
This schedule, therefore, prevents unstable guidance at early diffusion time while allowing the mutation-dependent force to influence the final data-like structure.

According to the theoretical formulation, DeltaDiff is expected to preferentially sample low-free-energy mutant conformations. In practice, however, the population of structures close to the experimentally or computationally determined mutant structures may still be limited, and in some cases only a small fraction of the sampled structures capture the expected mutant conformation.
For example, for the G10 mutant of Novispirin, the bent-helix structure can be found in the structures generated by DeltaDiff, but it is outnumbered by structures with smaller bending angles. This issue may arise from systematic errors in both the SBDM and the ML-CG model. Str2Str-generated protein structures are known to provide only a qualitative approximation to the Boltzmann distribution. Recently developed SBDMs, such as BioEmu \cite{lewis2025scalable}, provide improved approximations for selected proteins. More efforts are needed to develop SBDMs that can quantitatively model the Boltzmann distributions of proteins.
Other techniques, such as steered guidance \cite{10.5555/3692070.3694415,kulyte2024improving}, can also be applied to improve SBDM quality during inference. Such guidance strategies can be easily incorporated into DeltaDiff.
Similarly, errors in the ML-CG model can propagate to DeltaDiff. The ML-CG model used here was trained on only a small number of proteins and has not been broadly benchmarked for transferability across diverse protein systems. Nevertheless, our results show that this ML-CG model can still guide the SBDM to sample mutant-like structures. Recent developments in foundation CG models may further improve both accuracy and transferability \cite{charron2025navigating}. It is worth noting that the resolution of the CG model must match that of the SBDM. Since these foundation models typically use all backbone atoms and C$_\beta$ atoms as CG sites, an SBDM with the same CG degrees of freedom would be needed to directly incorporate them into DeltaDiff.
Because of the current limitations in available ML-CG models and SBDMs, DeltaDiff is best viewed as a structural exploration tool that can significantly increase the probability of sampling correct mutant conformations. However, as both ML-CG models and denoising-based foundation models continue to advance, the performance of DeltaDiff is expected to improve accordingly.

One major advantage of DeltaDiff, compared with conventional molecular dynamics simulation, is its sampling speed. The baseline model provides very fast protein structure generation. For the T8P chignolin case, Str2Str generated an independent structure with approximately \(0.36\) s per structure, while DeltaDiff uses approximately \(0.72\) s to generate an independent structure. 
This shows that DeltaDiff has computational efficiency close to that of the baseline model, with only a small additional cost from the MLCG guidance. 
Moreover, independent DeltaDiff can run in parallel on multiple GPUs. 
Therefore, DeltaDiff is expected to be suitable for high-throughput mutant-structure generation and mutant-design tasks. 
Even if DeltaDiff cannot uniquely determine the mutant structure, it is an efficient way to generate diverse structures that can be used in further relaxation and sampling \cite{10.1371/journal.pone.0301866, doi:10.1021/acs.jctc.2c01189, D5DD00201J}

\section{CONCLUSION}
In this work, we introduced DeltaDiff, a general framework that integrates mutation-dependent energy differences as physical guidance to capture non-local conformational changes induced by mutation. We tested this mutation-aware structure generation strategy with three representative systems: the T8P-associated turn rearrangement in chignolin, the bent helical geometry of Novispirin G-10, and the loop restructuring of BBL D162N. Across all three systems, DeltaDiff outperformed the baseline model by more effectively capturing key mutation-induced conformational changes.

One key strength of DeltaDiff is that mutation-aware structure sampling does not require fine-tuning or retraining of the baseline model. Instead of incorporating additional mutant structures into the training set, DeltaDiff introduces physical interaction guidance directly during inference. This provides an initial demonstration of how physics-guided inference can be used to capture mutation-induced structural changes.
Another important strength of DeltaDiff is its implementation flexibility. In this work, we used Str2Str as one example of the baseline model and an MLCG model as one example of physical guidance, demonstrating the ability of DeltaDiff to incorporate state-of-the-art SBDM and physical interaction models. As these models continue to improve, the performance of DeltaDiff can also be further enhanced. 

To summarize, DeltaDiff provides a new strategy that can potentially reduce the computational resources required to accurately capture nonlocal conformational changes induced by mutations, especially for systems involving a single mutation site, where current foundation models may struggle to distinguish mutant structures from wild-type structures based on sequence input alone. 

\begin{acknowledgement}

Y.C., Y.W., and M.C. gratefully acknowledge the support of the American Chemical Society
Petroleum Research Fund (Grant Number 67307). Computational resources were provided by Anvil at Purdue University through allocation CHE220008 from the Advanced Cyberinfrastructure Coordination Ecosystem: Services \& Support (ACCESS) program, which is supported by National Science Foundation Grants 2138259, 2138286, 2138307, 2137603, and 2138296.

\end{acknowledgement}

\bibliography{acs-achemso}

\end{document}